\documentclass[twocolumn,twoside,slac_two]{revtex4}
\usepackage{graphicx}
\usepackage{fancyhdr}
\begin{document}

\title{Super B Factories}

\author{M. Nakao}
\affiliation{
  KEK, High Energy Accelerator Research Organization, Tsukuba, Japan}

\begin{abstract}
After the establishment of the Kobayashi-Maskawa mechanism of $CP$
violation at the two $B$ factories, possibilities to increase the
integrated luminosity by two orders of magnitude have been investigated,
since it seems to be the amount needed to find physics beyond the
Standard Model through $CP$ violating and other observables in rare $B$
meson decays, $D$ meson decays and $\tau$ lepton decays.  This report
reviews the physics sensitivities and status of such super $B$ factories,
which are planned at two locations.
\end{abstract}

\maketitle

\thispagestyle{fancy}

                         \section{Introduction}

Confirmation of CP violation in $B$ meson decays through the measurement
of the time-dependent decay rate asymmetry has demonstrated the power of
the high luminosity $B$ factories, Belle at KEKB, KEK and BaBar at
PEP-II, SLAC.  Together with measurements of the angles and sides of the
Unitarity Triangle, the picture of CP violation through the
Kobayashi-Maskawa mechanism has been established to be at least the
dominant source of all CP violating phenomena in the high energy physics
known to date.  This was possible only by building the ``$B$ factories''
with two orders of magnitude higher luminosity than existing facilities
at that time, and with a new concept of the boosted center-of-mass frame
with asymmetric beam energies for an $e^+e^-$ collider.  The $B$
factories have also provided a rich field for many other non-trivial
tests of the Standard Model (SM) with their power based on a huge
statistics of $B$ decays, $D$ decays and $\tau$ decays.


However, in spite of its success, the SM is still considered to be a low
energy approximation of a more fundamental physics beyond the SM (BSM)
for many reasons. For example, CP violation in the SM cannot explain the
baryon number asymmetry in the universe; another more theoretical
argument is that BSM physics is expected to lie at the TeV energy scale
in order to solve the hierarchy problem. On the other hand, there is a
mystery called the flavor problem: a new interaction around a TeV energy
scale would alter the flavor changing neutral current amplitudes and a
large deviation due to BSM would be expected, unless the coupling
constants are fine tuned or aligned to the SM couplings to reconcile
with the strong constraints from $B$ factory measurements.

BSM effects should appear in most cases through loop diagrams, while
the tree diagrams are usually unaffected. To observe the effects on
the Unitarity Triangle of different contributions of loop and tree
diagrams, and also through other rare phenomena, it is again necessary
to build a new facility with a further two orders of magnitude
higher luminosity than now. This is the motivation for a ``super'' $B$
factory.


\def\calS{{\cal S}}
\def\calA{{\cal A}}
\def\Vub{V_{ub}}
\def\Vcb{V_{cb}}
\def\Vtb{V_{tb}}
\def\Vtd{V_{td}}
\def\ubar{\overline{u}{}}
\def\cbar{\overline{c}{}}
\def\qbar{\overline{q}{}}
\def\btoc{b\to c}
\def\btou{b\to u}
\def\btos{b\to s}
\def\btod{b\to d}
\def\btoccs{b\to c\cbar s}
\def\btosqq{b\to s\qbar q}
\def\btoclnu{b\to c\ell\nu}
\def\btoulnu{b\to u\ell\nu}
\def\BtoXulnu{B\to X_u\ell\nu}
\def\Btopilnu{B\to \pi\ell\nu}
\def\piPM{\pi^\pm}
\def\piP{\pi^+}
\def\piM{\pi^-}
\def\piZ{\pi^0}
\def\KZ{K^0}
\def\KS{K^0_S}
\def\KL{K^0_L}
\def\Jpsi{J/\psi}
\def\DZ{D^0}
\def\Dbar{\overline{D}{}}
\def\DZbar{\Dbar^0}
\def\DZDZbar{\DZ\DZbar}
\def\BZ{B^0}
\def\Bbar{\overline{B}{}}
\def\BZBZbar{\BZ\Bbar^0}
\def\BtoJpsiKS{B\to\Jpsi\KS}
\def\BtopiPpiMpiZ{B\to\piP\piM\piZ}
\def\BtopiPpiM{B\to\piP\piM}
\def\Btopipi{B\to\pi\pi}
\def\Btorhopi{B\to\rho\pi}
\def\Btorhorho{B\to\rho\rho}
\def\BtophiKZ{B\to\phi\KZ}
\def\BtoetapKZ{B\to\eta'\KZ}
\def\BtoKSKSKS{B\to\KS\KS\KS}
\def\Btotaunu{B\to\tau\nu}
\def\BtoDorDstartaunu{B\to D^{(*)}\tau\nu}
\def\Btomunu{B\to\mu\nu}
\def\BtoDZK{B\to D^0K}
\def\abinv{\;{\rm ab}^{-1}}
\def\lumiunit{{\rm cm}^{-2}{\rm s}^{-1}}
\def\EM#1{\times10^{-#1}}
\def\EP#1{\times10^{#1}}

                        \section{Physics Programs}

The key measurement at a super $B$ factory will still be the
measurement of the Unitarity Triangle, but with a precision an order
of magnitude better than what we have now. There are also a large
number of other potential measurements that are sensitive to BSM, and
their importance will be more significant than has been the case in
Belle and BaBar. More thorough and detailed studies can be found in
\cite{superkekb-loi,superb-cdr,sff,yellow-report}.

\subsection{Unitarity Triangle}

The current world average on the angle $\phi_1$ of the Unitarity
Triangle\footnote{In this report, the notation $\phi_1$, $\phi_2$,
  $\phi_3$ is used instead of $\beta$, $\alpha$ and $\gamma$.} has a
remarkable precision, $\phi_1=(21.1\pm0.9)^\circ$~\cite{hfag2008},
where the error is dominated by the results from the two $B$ factories
using the sum of the datasets exceeding $1\abinv$. Measurement of this
and other angles, and the sides of the Triangle were made possible
only after the $B$ factories came online. All the results so far are
found to be consistent with each other. However, they are not precise
enough yet to identify any possible discrepancy, for example in the
$CP$ violating phase measured in the loop diagrams in which BSM
effects could reside, from the phase measured from tree diagrams in
which BSM effects are not expected.

The angle $\phi_1$ is measured in the form of $\sin2\phi_1$, which is
the coefficient of the sine term ($\calS$) in the time-dependent $CP$
asymmetry that appears due to an irreducible complex phase in the
$\BZBZbar$ mixing when measured together with the $\btoccs$ transition
into a $CP$ eigenstate such as $\Jpsi\KS$.  At this moment the measurement
of $\sin2\phi_1$ is still slightly statistical error dominated.  An
early dataset of $5\abinv$ from a super $B$ factory will turn this into
an ultimate measurement of $\sin2\phi_1$ with an error of $0.015$ (or
$0.6^\circ$ in terms of $\phi_1$).  The precision of the measurement
then matches the theoretical uncertainty, which appears due to the
effects of subdominant diagrams with the same final state, a complex
phase different from $\phi_1$ and an unknown size of the amplitude.

The angle $\phi_2$ appears in the combination of the $\BZBZbar$ mixing
and the $\btou$ transition into a $CP$ eigenstate such as
$\BtopiPpiM$. Unfortunately, $\sin2\phi_2$ cannot be directly measured
from the time-dependent asymmetry of $\BtopiPpiM$ due to the $\btod$
penguin contribution which has a different weak phase. The amplitude
of such a contribution and the angle $\phi_2$ have to be extracted
from measurements of all $\pi\pi$ charge combinations ($\piP\piM$,
$\piPM\piZ$, $\piZ\piZ$) by assuming isospin symmetry. The same is
possible with $\Btorhorho$, or a more elaborate technique can be
performed to disentangle the amplitudes and phases of $\Btorhopi$
using a time-dependent Dalitz analysis of $\BtopiPpiMpiZ$. Multiple
solutions appear in these procedures and they make it difficult to
pinpoint the value of $\phi_2$; the current $1\sigma$ interval from
combined Belle and BaBar results is $[83.5, 94.0]^\circ$, e.g.,
in~\cite{ckmfitter}. A super $B$ factory will significantly improve
the situation as the correct solution will become unambiguous from the
three independent measurements. The combined error for $\phi_2$ is
about $2^\circ$ with $5\abinv$ data. This is the ultimate precision at
which the experimental error matches the model uncertainty of the
Dalitz amplitudes or theoretical uncertainty due to isospin breaking
effects.


At a super $B$ factory, the set of angles $\phi_1$ and $\phi_2$,
already with $5\abinv$, provides a reference ``point'' of the
Unitarity Triangle with a 2\% error. All of the other possible
measurements will be used to search for a deviation of $O(10\%)$ from
the SM.

\subsection{Deviation from the Unitarity Triangle}

The three most promising measurements to search for a deviation from
expectation for the Unitarity Triangle variables are; the angle
$\phi_1$ from loop mediated $\btos$ decays, the angle $\phi_3$ from
the tree process $\BtoDZK$, and the length of the side $|\Vub|$ using
$\btoulnu$ decays. Any deviation from the reference point defined by
$(\phi_1, \phi_2)$ becomes evidence of BSM.

\begin{figure}[h]
\centering
\includegraphics[width=80mm]{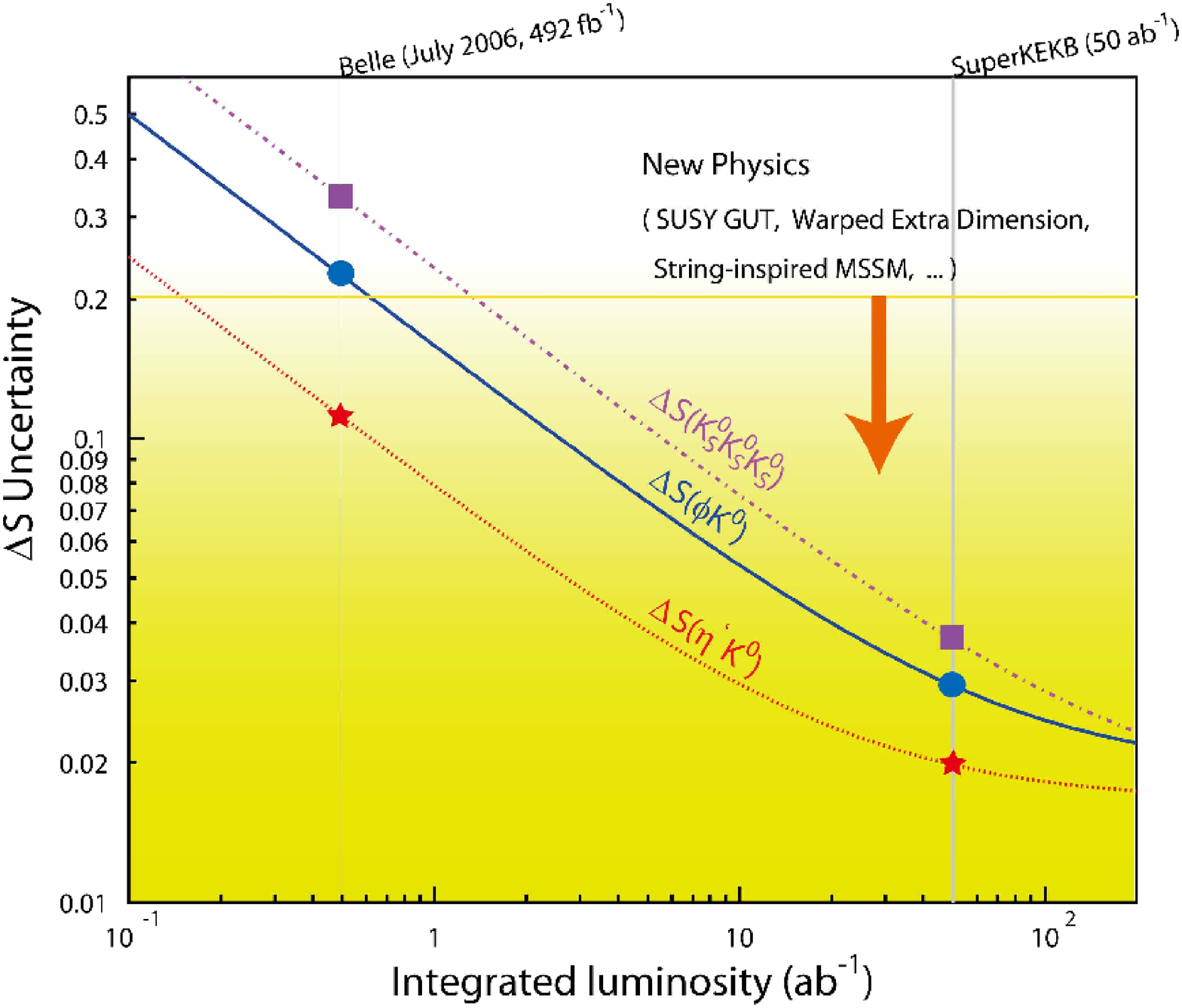}
\caption{Expected sensitivities of the CP violation measurements in
  $\BtoetapKZ$, $\BtophiKZ$ and $\BtoKSKSKS$ decays as a function of the
  integrated luminosity at SuperKEKB.  } \label{fig:btoscpv}
\end{figure}

In the SM, the size of the time dependent $CP$ asymmetry $\calS$ in
the $\btosqq$ decay modes into a $CP$ eigenstate is the same as that
in $\btoccs$, i.e.\ $\sin2\phi_1$, as both diagrams have the same weak
phase. Since $\btosqq$ is a penguin loop, the phase can be modified by
an additional BSM amplitude if it exists, while it is unchanged in
$\btoccs$. Therefore, the deviation in $\calS$ from $\sin2\phi_1$ is
the sign of BSM. For a given final state, there are always subdominant
diagrams, e.g., the tree diagram $\btou$ with $s\ubar$, which are the
source of possible modification of $\calS$ within the SM. Three modes,
$\BtophiKZ$, $\BtoetapKZ$ and $\BtoKSKSKS$, are considered to be the
golden modes with the smallest of such pollutions, estimated to be
about 0.02. Measurements will be statistical error dominated until
$50\abinv$ or more integrated luminosity is obtained, as shown in
Fig.~\ref{fig:btoscpv}. The expected errors are, 0.02, 0.03 and 0.04
for $\BtoetapKZ$, $\BtophiKZ$ and $\BtoKSKSKS$, respectively, at
$50\abinv$.

The angle $\phi_3$ is most cleanly measured using $\BtoDZK$ decays.
There, $\phi_3$ appears as interference between the tree amplitude
$\btoc$ with $d\ubar$ and another tree amplitude $\btou$ with $d\cbar$.
The former gives $\DZbar$ in the final state while the latter gives
$\DZ$; common final states between $\DZ$ and $\DZbar$ decays are the
source of interference between these two decay channels.  As there is no
subdominant amplitude with a loop diagram, this provides the cleanest SM
measurement.  There are several methods depending on different $\DZ$
decay modes, e.g., $\DZ$ decaying into a $CP$ eigenstate, or into a doubly
Cabibbo suppressed final state.  The most effective method is based on
the Dalitz analysis of the decay chain $B^\pm\to D K^\pm$; $D\to
\KS\piP\piM$.  The analysis depends on the model of the three-body $D$
decay amplitudes, and gains significantly from charm factory data with
CP-tagged $\DZ$ decays.  The combined $\phi_3$ error will be $6^\circ$
at $5\abinv$, or $2^\circ$ at $50\abinv$.


The $|\Vub|$ measurement is performed using either an inclusive
measurement of $\BtoXulnu$ or an exclusive one such as $\Btopilnu$.
Currently, both measurements have a similar size in the error, and
have some tension between the results. Both methods provide the
cleanest results using a tagging technique of reconstructing a
hadronic $B$ decay mode for the other $B$ meson decay. The tagging
efficiency is not very high and demands huge statistics which is
suitable at a super $B$ factory. Since only a limited kinematical
range can be measured due to the huge $\btoclnu$ background, the
inclusive measurement requires the operator product expansion
technique to extrapolate into the full kinematical range. All the
necessary information is available from data, and the total error will
be 6\% at $5\abinv$ and 4\% at $50\abinv$. The exclusive branching
fraction will be more accurately measured at a super $B$ factory. If
the $B\to\pi$ form factor is calculated using lattice QCD more
precisely, $|\Vub|$ will be determined from the exclusive measurement
with a smaller error.

\begin{figure}[h]
\centering
\includegraphics[width=80mm]{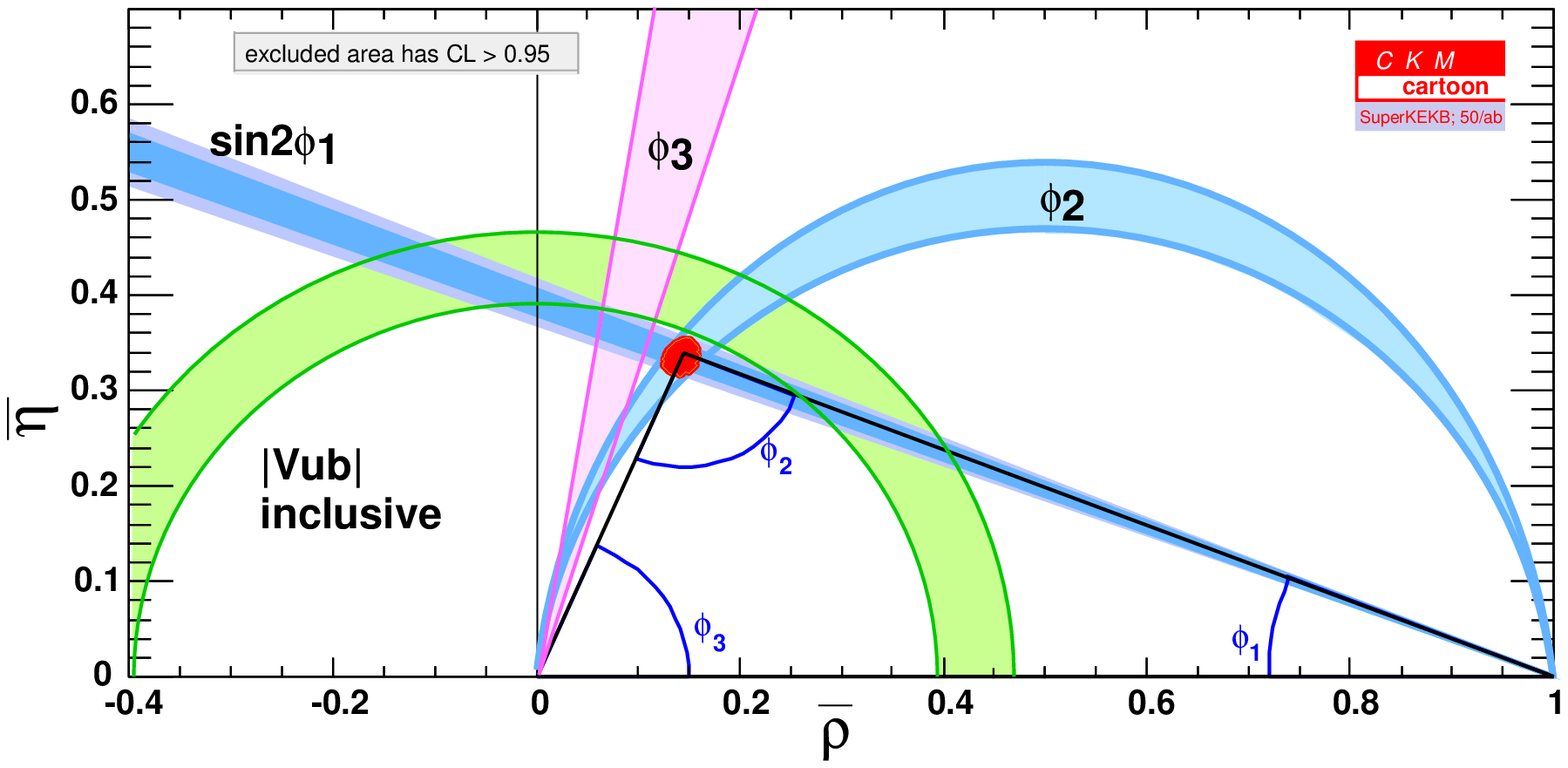}
\caption{Expected sensitivities of the Unitarity Triangle measurements
  at $50\abinv$ data at SuperKEKB.
  } \label{fig:rhoeta50}
  
\end{figure}

Expected sensitivities of the Unitarity Triangle measurements at
$50\abinv$ data at SuperKEKB is shown in Fig.~\ref{fig:rhoeta50} using
the current world average values for the central values.
Given these measurements, a 10\% deviation in any of these measurement
would be identified at a super $B$ factory with $50\abinv$.

\subsection{More Key Measurements}

There are many other key measurements which can be used to search for
BSM in $B$ decays, and also in decays of $D$ mesons and $\tau$ leptons,
which are equally abundantly produced at a super $B$ factory.

The weak interaction which governs $b$ quark decays is based on a
left-handed current in the limit of massless quarks. This is not
necessarily the case in many BSM models. The right-handed current in
the $\btos$ transition can be effectively identified as a non-zero
$\calS$ value in a time-dependent CP measurement of
$B\to\KS\piZ\gamma$. With $50\abinv$ the error will be less than $3\%$
and already better than the theoretical uncertainty on the deviation
of $\calS$ from zero due to the finite mass of the $s$ quark.

Many BSM models also require more than one Higgs doublet, including a
charged Higgs boson. The charged Higgs boson can replace the weak
boson in a tree diagram, and its effect is enhanced in the helicity
suppressed purely leptonic decays and semi-leptonic decays with a
$\tau$ lepton. The effect can be searched for through a deviation from
expectation in the branching fraction of $\Btotaunu$. Similar
measurements can be performed with $\BtoDorDstartaunu$ and $\Btomunu$.
If deviations are observed in all these modes, a comparison between
them leads to a test of the universality of the coupling, and provides
stronger evidence for the existence of the charged Higgs boson.

Inclusive measurements such as $B\to X_s\gamma$, $B\to X_d\gamma$ and
$B\to X_s\ell^+\ell^-$ are also sensitive to a wide range of BSM.
Especially, the zero-crossing point of the forward-backward asymmetry in
$B\to X_s\ell^+\ell^-$ has a very clean signature.

The recently observed large values of the $\DZDZbar$ mixing parameters
$(x,y)$, of the order $10^{-2}$, suggest the possibility of a BSM
contribution, while an explanation within the SM is not excluded because
of a large hadronic uncertainty.  A measurement of $CP$ violation in
$\DZDZbar$ mixing would be clear evidence for a BSM effect in the charm
quark sector.

Finally, lepton flavor violating $\tau$ decay is also allowed in many
BSM models, while it is not allowed at all in the SM. There are a
large number of possible lepton flavor violating decay modes (e.g.,
$\tau\to\mu\gamma$, $\tau\to\mu \eta$ or $\tau\to e^+e^-e^+$) which
have been and will be searched for. If observed, it will be an
unambiguous sign of new physics.

\subsection{Comparison with LHCb}

There may be a question why we have to build a super $B$ factory while
the next generation flavor physics can be studied at LHCb. In reality,
it is almost impossible to measure modes with photons, $\piZ$ and
neutrinos, and perform inclusive measurements at LHCb. Many of
these are the key measurements to study BSM as already discussed.

There are examples where LHCb has an excellent sensitivity: the
Unitarity Triangle parameters, especially the angle $\phi_3$, can be
precisely measured at LHCb with a similar precision to that of at a
super $B$ factory, provided that the systematic errors are under
control. In order to search for a BSM $CP$ phase in the $b\to s$
transition, $B_s\to\phi\phi$ can be used; in order to search for the
right-handed current, $B_s\to\phi\gamma$ can be used. These are
different decay modes related to searches for the same type of BSM
effects, and the searches at the two places are extremely helpful for
gaining an unambiguous understanding of BSM physics.

\section{Next Generation B Factories}

In order to collect an integrated luminosity of $50\abinv$ within a
reasonable amount of running time, the instantaneous luminosity has to
be above or at least close to $10^{36}\lumiunit$. In addition, to keep
synergy with energy frontier physics at the LHC and flavor physics at
LHCb, it is crucial to operate the super $B$ factory in the next
decade.

Currently, two projects are planned: the SuperKEKB project in Japan and
the SuperB project in Italy.  If resources allow, it is definitely
better to have both facilities for healthy competition and cross-checks,
as was extremely helpful in the case of competition between Belle
and BaBar.  However, under the current situation for high energy
physics, it does not seem to be possible to have both of them at the
same time.

Key parameters for a high luminosity are the beam current ($I$) and the
beam-beam parameter ($\xi_y$) that are proportional to the instantaneous
luminosity, and the vertical $\beta$ function at the interaction point
($\beta_y^*$) which is inversely proportional to the luminosity.  The
two projects take different approaches for a higher luminosity.  Typical
parameters for SuperKEKB are beam currents of 9.4 A $\times$ 4.1 A,
$\xi_y>0.24$ and $\beta_y=3.0$ mm for both rings.  Those for SuperB are
1.85~A $\times$ 1.85~A beam currents, $\xi_y=0.15$ and
$\beta_y=0.39/0.22$ mm for the high/low energy ring.  An example of the
parameters is given in Table~\ref{tbl:superkekb-superb}.  These
parameters give instantaneous luminosities of $0.8\EP{36}\lumiunit$ for
SuperKEKB and $1.0\EP{36}\lumiunit$ for SuperB, almost two orders of
magnitude higher than the current KEKB record, $0.017\EP{36}\lumiunit$.

\def\twoc#1{\multicolumn{2}{c}{#1}}

\begin{table}[h]
\begin{center}
\caption{An Example of machine parameters for SuperKEKB and SuperB.}
\label{tbl:superkekb-superb}
\begin{tabular}{lcccccc}
    \hline
    & & \twoc{SuperKEKB} & & \twoc{SuperB} \\
    \hline
    && ($e^+$) & ($e^-$) && ($e^+$) & ($e^-$) \\
    Energy (GeV)      && 3.5 & 8     && 4 & 7 \\
    Luminosity
          ($10^{36}$) && \twoc{0.55}  && \twoc{1.0} \\
    Number of bunches && \twoc{5018} && \twoc{1251} \\
    Beam current (A)  && 9.4 & 4.1   && 1.85 & 1.85 \\
    $\beta(y^*)$ (mm) &&  \twoc{3}        && 0.22 & 0.39 \\
    $\beta(x^*)$ (mm) &&  \twoc{200}      && 35 & 20 \\
    emittance $\epsilon(y)$ (pm.rad) && 60 & 66 && 7   & 4   \\
    emittance $\epsilon(x)$ (nm.rad) && 12 & 13 && 2.8 & 1.6 \\
    beam-size $\sigma(x^*)$ ($\mu$m) && 37.5 & 39.8 && 0.039 & 0.039 \\
    beam-size $\sigma(y^*)$ ($\mu$m) && 2.11 & 2.28 && 9.9 & 5.66 \\
    bunch length           && \twoc{3}    &&     &      \\
    Damping time (ms) && 84/-- & 47/-- && 40/20 & 40/20 \\
    Touschek lifetime (min)        &&   &   && 20 & 40 \\
    Beam lifetime (min)  &&   &   && 5.0 & 5.7 \\
    tune-shift $\xi(y)$ && \twoc{0.296} && \twoc{0.15} \\
    tune-shift $\xi(x)$ && \twoc{0.153} && 0.0043 & 0.0025 \\
    RF power (MW)            && & && \twoc{17} \\
    \hline
  \end{tabular}
\end{center}
\end{table}

\subsection{SuperKEKB}

The SuperKEKB project is an upgrade of the current KEKB facility at the
same place, reusing a large fraction of the existing components and
infrastructure.

One of the key components of SuperKEKB is the ``crab'' cavity that
rotates the envelope of the beam bunch and makes head-on collisions
possible for beams incident at a finite angle (crab-crossing). This will
at least geometrically increase the effective volume of the collision.
According to a simulation, the effect is more dramatic: the beam-beam
force becomes nearly independent of the horizontal coordinate at a half
integer tune. In the case of KEKB with a 22 mrad crossing angle, $\xi_y$
becomes 0.15 and almost doubles the luminosity, and for SuperKEKB with a
30 mrad crossing angle, $\xi_y>0.24$ is possible.

\begin{figure}[h]
\centering
\includegraphics[width=80mm]{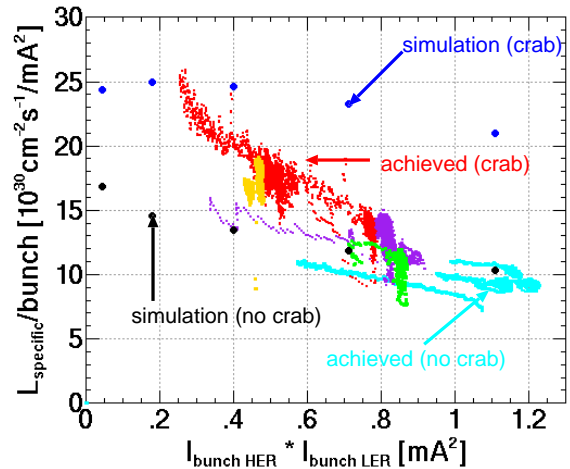}
\caption{Specific luminosity simulated and measured with and without the
crab cavity at KEKB.} \label{fig:speclumi-crab}
\end{figure}

With this simulation result, KEKB has installed a crab cavity for each
of high and low energy rings and has been commissioning since 2007.
Under a low beam current operation, the specific luminosity reached
the predicted value (Fig.~\ref{fig:speclumi-crab}), and an enhancement
in the beam-beam parameter, $\xi_y=0.092$, with respect to the case
before the crab cavity, $\xi_y=0.056$, was observed. However, there is
a drop in the luminosity under a high beam current, and the reason is
still being investigated. In addition, the current crab cavity has to
be operated at a lower beam current than what was already achieved
without, and this has prevented the expected boost in the
instantaneous luminosity so far. The commissioning will continue for
one more year until the end of the KEKB running time.

The other key issue is the higher beam current. In order to store a
higher beam current without being affected by the electron cloud, the
vacuum pipe will be replaced all over the ring with antechamber type
beampipes, and the bellows with higher-current-proof ones.

The KEKB upgrade plan is already included in KEK's five-year roadmap
from 2009 to 2013. This includes a three-year shutdown time for the
construction.

\subsection{Italian SuperB}

The SuperB project in Italy is a completely new project, with an ultra
low emittance approach inspired by the International Linear Collider
(ILC) project.  In order to achieve a very high luminosity with a
moderate beam current, both beams have to be squeezed into an extremely
small interaction region with a vertical size of 39 nm, and with a very
small $\beta_y^*$.  Such a beam could be possible, according to the
studies for the ILC dumping ring.  However, it does not necessary mean
that such a small emittance can be sustained while making collisions at
every turn with a very high luminosity.  The other problem would be the
very short beam lifetime of about only five minutes.

\begin{figure}[h]
\centering
\includegraphics[width=80mm]{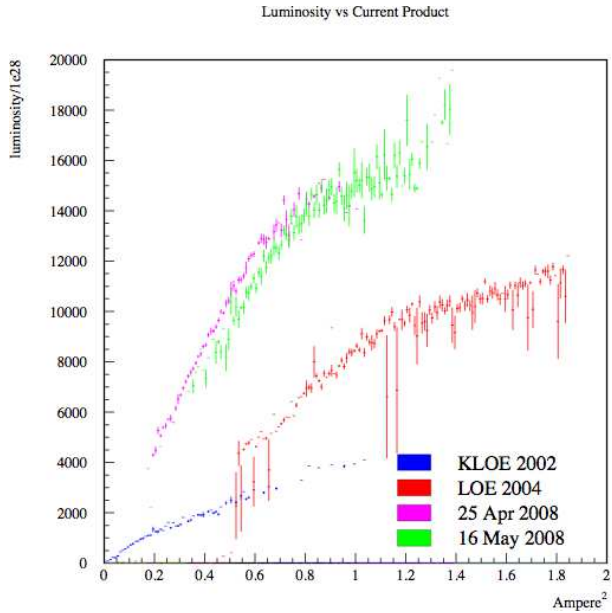}
\caption{Luminosity improvement with the crab waist scheme measured at
  DA$\Phi$NE.} \label{fig:dafne-2008may}
\end{figure}

The ultra-small beamsize also implies a huge hourglass effect, and it
has to be suppressed either by making the bunch length extremely
short, which is not realistic, or by moving the vertical waist
positions of both beams along the $z$ axis with a proper phase. The
latter is called the ``crab waist'', and can be realized with a
sextupole magnet. The concept of the crab waist has been successfully
tested at DA$\Phi$NE at a low energy and low beam current, as shown in
Fig.~\ref{fig:dafne-2008may}.

The other consideration at SuperB is to make the energy asymmetry
smaller, in order to save on the electricity bill. This makes the
resolution of the time difference between two $B$ meson vertices
worse, but it gives a better hermeticity, which is crucial for
measurements with neutrinos.

The candidate site is at the Tor Vegarta campus of INFN Rome.  Here
also, many components, including magnets, will be brought from PEP-II to
minimize the cost.

\section{Detector Considerations}

A high luminosity brings a high event rate. The $B$-pair events alone
will be delivered at a rate of 1 kHz under a luminosity of
$1\EP{36}\lumiunit$, and the physics trigger rate will become 10 kHz
excluding the Bhabha events, which have an even higher rate. In order
to keep nearly 100\% efficiency for the $B$-pair events, the trigger
and data acquisition system have to be drastically improved. The
readout electronics also has to be as deadtime-free as possible.

A high luminosity brings a high background rate at the same time.  On
the other hand, to keep the advantage of the clean $e^+e^-$ environment
especially for measurements with photons, $\piZ$ and neutrinos, any
performance drop with respect to the current Belle or BaBar detector
is not acceptable.

Since the existing Belle and BaBar detectors already have excellent
performance, it is not easy to drastically improve it, especially after
coping with the high background.

\subsection{SuperKEKB Detector}

The detector at SuperKEKB (SuperBelle) has to cope with the large
beam-gas background due to the much higher beam currents. After a
careful design of the beampipe and masks at the interaction region, it
is found that the size of Touschek background is moderate, and the
radiative-Bhabha background is not harmful except for the outmost
$K_L$ and muon detector. The total background will be about 20 times
that of the current conditions at Belle, and therefore a significant
amount of modification to the Belle detector is necessary.

\begin{figure}[h]
\centering
\includegraphics[width=80mm]{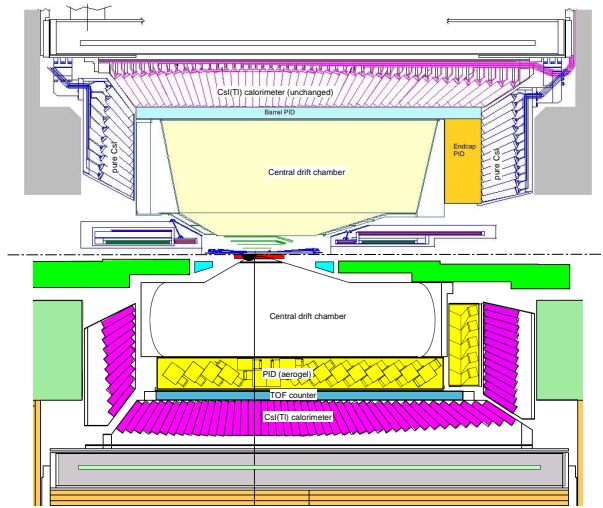}
\caption{Comparison between the SuperBelle detector (upper) and the
  Belle detector (lower).} \label{fig:superbelle}
\vspace{12pt} 
\end{figure}

\begin{figure}[h]
\centering
\includegraphics[width=80mm]{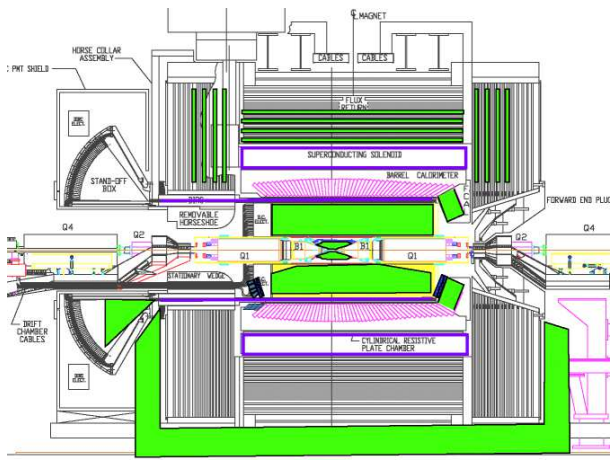}
\caption{SuperB detector without (upper) and with (lower) optional
  detector components.} \label{fig:superb-det}
\end{figure}

In order to cope with the background, the following changes are
planned. The silicon vertex detector will have a larger radius with
six layers, with a possible option of a pixel detector for the
innermost layer. The enlarged vertex detector will replace the inner
part of the drift chamber and allow a larger volume for
$\KS\to\piP\piM$ vertexing. The drift chamber will be replaced with
one with smaller cell size to shorten the drift time and to reduce the
occupancy. The outer radius of the drift chamber will be enlarged,
thanks to the thinner outer detectors. The particle identification
devices will be fully replaced from existing time-of-flight counters
and the threshold-type aerogel Cherenkov counter, with a detector to
reconstruct the Cherenkov ring image, such as a time-of-propagation
counter for the barrel part and an aerogel ring-image counter for the
forward endcap part. The endcap part of the calorimeter will be
replaced with pure CsI crystals that have a faster time response,
while the thallium-doped CsI crystals will be unchanged for the barrel
part. The resistive plate counters for $\KL$ and muon detection will
be replaced with scintillation fibres.

\subsection{SuperB Detector}

Thanks to the smaller beam current, the beam-gas background is expected
to be moderate at the SuperB detector.  On the other hand, a huge
Touschek background is expected, and also the very short lifetime could
be harmful for the detector.

The detector is based on reuse of the existing BaBar detector, which
is already more immune to backgrounds than Belle.  However, similarly to
Belle's case, many of the components have to be replaced.  These
includes a new silicon vertex tracker, a new drift chamber, a new
forward calorimeter and a new $\KL$ and muon detection system.

\vspace*{12pt} 

\section{Summary}

The physics program at a super $B$ factory is very compelling in the
integrated luminosity range between 5 to 50 $\abinv$.  This includes
the measurements of the precise reference point in the Unitarity
Triangle and possible deviations from there, extensive searches for
right handed currents, charged Higgs, lepton-flavor violating decays, and
many other search channels.

Both projects described in this paper are actively working on the
accelerator and detector designs. The design for SuperKEKB is already
finalizing for the production of necessary components, while the
SuperB design is also getting converged. We are looking forward to the
exciting future of flavor physics that will be possible at a super $B$
factory.


\bigskip 

\end{document}